\documentclass[12pt]{article}
\usepackage{psfig,epsf}
\usepackage{a4wide,graphicx}
\usepackage{cite}

\begin{document}

\title{Sequential vector and axial-vector meson exchange and chiral loops in
radiative phi decay}

\author{
J.~E.~Palomar, L.~Roca, E.~Oset and M.~J.~Vicente Vacas\\
{\small Departamento de F\'{\i}sica Te\'orica and IFIC,
Centro Mixto Universidad de Valencia-CSIC,} \\ 
{\small Institutos de
Investigaci\'on de Paterna, Aptdo. 22085, 46071 Valencia, Spain}\\ 
}

\date{\today}

\maketitle \begin{abstract} We study the radiative $\phi$
decay into $\pi^0 \pi^0 \gamma$ and  $\pi^0 \eta \gamma $
taking into account mechanisms in which there are two
sequential vector-vector-pseudoscalar  or
axial-vector--vector--pseudoscalar steps followed by the
coupling of a vector meson to the photon, considering the
final state interaction of the two mesons.  There are other
mechanisms in which two  kaons are produced through the same
sequential mechanisms or from $\phi$ decay into two kaons and
then undergo final state interaction leading to the final
pair of pions or $\pi^0 \eta$,
this latter mechanism being the leading one. 
The results of the parameter free theory, together with the
theoretical uncertainties, are compared with the latest
experimental results of KLOE at Frascati. 

\end{abstract}

\section{Introduction}

The radiative decays of the $\phi$ into $\pi^0 \pi^0 \gamma$ and 
$\pi^0 \eta \gamma$ have been the subject of intense study
\cite{greco,franzini,colangelo,achasov,bramon5,lucio,uge}. 
One of the main reasons for this is the hope that one can get much information
about the nature of the $f_0(980)$ and $a_0(980)$  resonances 
from the invariant mass distribution of the two mesons.  The nature of the scalar meson 
resonances has generated a large debate  \cite{kyoto}, to which new 
light has been brought by the 
claim that  these resonances are dynamically generated from multiple
scattering with the ordinary chiral
Lagrangians  \cite{npa,iam,nsd}.

   These two reactions involving the decay of the $\phi$ are 
special. Indeed, the $\phi$ does not decay into two pions
because of isospin symmetry. 
  A way to
circumvent this handicap is to allow the $\phi$ to decay into
two charged kaons (with a photon attached to one of them) and
the two kaons to scatter giving rise to the two pions (or
$\pi^0 \eta$).  The loop which appears diagrammatically is
proved to be finite using  arguments of gauge invariance 
\cite{pestieau,close,bramon5}.  The radiative $\phi$ decay
through this mechanism was studied in \cite{bramon5} and the
results of lowest order chiral perturbation theory ($\chi
PT$) were used to account for the $K^+ K^- \to \pi^0 \pi^0$
transition.  Since the  chiral perturbation theory $K^+ K^-
\to \pi^0 \pi^0$ amplitude does not account for the
$f_0(980)$, the excitation of this resonance has to be taken
in addition, something that has been done more recently using
a linear sigma-model in \cite{bramonnew}.  

  The work of \cite{uge} leads to the excitation of the 
$f_0(980)$  in the  $\pi^0 \pi^0$ production, or the
$a_0(980)$ in $\pi^0 \eta$ production in a natural way, since
the use of unitarized chiral perturbation theory ($U \chi
PT$), as in \cite{npa}, generates automatically those
resonances in the meson meson scattering amplitudes and one
does not have to introduce them by hand. 

The experimental situation has also experienced an impressive
progress recently.  To the already statistically significant 
experiments at Novosibirsk \cite{novo1,novo2,novo3} one has
added the new, statistically richer,
experiments at Frascati
\cite{frascati1,frascati2} which allow one to test models
beyond the qualitative level.   In this sense although
the predictions of the work of \cite{uge}, using $U \chi PT$
with no free parameters, provided a good agreement with the
experimental data of \cite{novo1,novo2,novo3}, thus settling
the dominant mechanism as that coming from chiral kaon loops
from the $\phi \to K^+ K^-$ decay, the new and more  precise
data leave room for finer details which we evaluate in this
paper.

 One of the issues concerning radiative decays is the
relevance of  the sequential vector meson mechanisms.  The
anomalous vector-vector-pseudoscalar couplings
\cite{singer,grau}, followed by the coupling of the photon to
the vector mesons through vector meson dominance, allows
mechanisms for reactions of the type that we study, through a
sequential $V \to VP \to PP \gamma$ process. This mechanism
is known to provide the $\omega \to \pi^0 \pi^0 \gamma$
radiative width with accuracy \cite{grau} and has been
further extended to study  $\rho \to \pi^0 \pi^0 \gamma$ and
other radiative decays in \cite{escribano,palomar}.  In
particular, it was found that in the latter  reaction the
sequential vector meson mechanism  and the  loops (in this
case  charged pion loops) were equally important, and the
consideration of the two mechanisms produced a width
compatible with the recent experimental determination of the
SND collaboration \cite{snd}.  Since the $\phi \to \rho
\pi^0$ is OZI forbidden, the contribution from the sequential
vector meson mechanism should be small and it was not
considered till recently \cite{achasovlast,lucio}. It
proceeds via $\phi - \omega$ mixing, which is indeed small,
but still noticeable in the
radiative $\phi$ decay width.  

  In the present paper we also take into account this
sequential mechanism, but in addition we consider the final
state interaction of the two mesons
within the framework of $U\chi PT$.
  Furthermore,  in the
line of considering the final state interaction (FSI) of the mesons
we also allow the sequential vector meson production of
kaons, which interact among themselves and lead to the final
two mesons.  This is easily  implemented in the  coupled
channel formalism of \cite{npa}.  The mechanisms involving
final state interaction all contribute to the production of
the $f_0(980)$ or $a_0(980)$ resonances, since the resonance
poles appear in the coupled channel formalism  in all the
elastic or transition matrix elements.

  Another novelty of the present work is the  consideration
of sequential mechanisms involving the exchange of an
intermediate axial-vector meson ($J^{PC}=1^{++}$ or
$1^{+-}$), both producing directly the final meson pair or
through the intermediate production of kaons which undergo
collisions and produce these mesons.

   All the mechanisms considered here contribute moderately,
but appreciably, to the $\phi$ radiative width. The inclusion
of all these mechanisms leads to results compatible with the
experimental data of Frascati, particularly if theoretical
uncertainties are considered, which is something also done in
the present work.

   The good agreement with experiment is reached in spite of
having in our approach a width for the $f_0(980)$ very small,
of the order of 30 MeV, seemingly in contradiction with the
'visual'  $f_0(980)$ width in the experiment, which looks
much larger. The reason for this has been recently discussed
in  \cite{achasovq,pennington} and stems from the fact that,
due to gauge invariance, the amplitude for the process
contains as a factor the momentum of the photon, which grows
fast as we move down to smaller invariant masses
of the two pseudoscalars
 from the
mass of the  $f_0(980)$ where the photon momentum is very
small.  This distorts the shape of the resonance, making it
appear wider.  We shall see that our approach, which respects
gauge invariance, introduces automatically this factor in the
amplitudes.

   We shall see that there is some discrepancy of the
theoretical results  with the data at small two meson 
invariant masses.  We shall discuss this feature, realizing
that the results resemble very much the raw data, before the
analysis is done to subtract the contribution of
$\omega\pi^0$ and to correct for the experimental acceptance.
Furthermore, some of the assumptions made in the analysis of 
\cite{frascati1} might be  questionable.

\section{The $\phi\to\pi^0\pi^0\gamma$ decay}

\subsection{Kaon loops from $\phi \to K^+ K^-$ decay
\label{sec:chiral_loops}}

\begin{figure}
\centerline{\protect\hbox{
\psfig{file=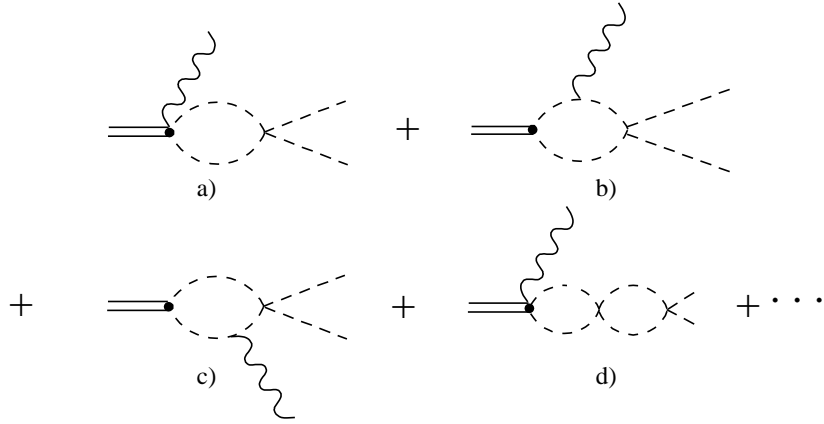,width=0.7\textwidth,silent=}}}
\caption{Loop diagrams included in the chiral loop contributions. The
intermediate states in the loops are $K^+K^-$. }
\label{looplot}
\end{figure}

  The mechanism for radiative decay using the tensor formulation for the vector
mesons has been discussed in \cite{neufeld,uge} and we briefly summarize it
here. The diagrams considered are depicted in Fig.~\ref{looplot}, where the 
loops contain 
$K^+ K^ -$. The vertices needed for the diagrams are obtained from
the chiral Lagrangian

\begin{center}
\begin{equation}   \label{eq:LFVGV}
 {\cal L} = \frac{F_V}{2 \sqrt{2}} < V_{\mu\nu} f^{\mu\nu}_{+} >
+ \frac{iG_V}{\sqrt{2}} < V_{\mu\nu} u^{\mu}u^{\nu} > ,
\label{resolagr}
\end{equation}
\end{center}
where the notation is defined in \cite{ecker}. We also assume ideal
mixing between the $\phi$ and the $\omega$ 

\begin{equation}
\sqrt{\frac{2}{3}} \omega_1 + \frac{1}{\sqrt{3}} \omega_8 \equiv \omega
\quad , \qquad
  \frac{1}{\sqrt{3}} \omega_1 - \frac{2}{\sqrt{6}} \omega_8 \equiv \phi
\end{equation}

\noindent
and the matrix $V$ is then given by

\begin{equation}
V_{\mu \nu} \equiv \left(\begin{array}{ccc} 
\frac{1}{\sqrt{2}}\rho^0_{\mu \nu} + \frac{1}{\sqrt{2}} \omega_{\mu \nu}
 & \rho^+_{\mu \nu} & K^{* +}_{\mu \nu} \\
\rho^-_{\mu \nu}& -\frac{1}{\sqrt{2}} \rho^0_{\mu \nu}
+ \frac{1}{\sqrt{2}} \omega_{\mu \nu} & K^{* 0}_{\mu \nu} \\
K^{* -}_{\mu \nu} & \bar{K}^{*0}_{\mu \nu} & \phi_{\mu \nu}
\end{array}
\right)
\end{equation}

The amplitude for the radiative decay can be written as
$T^{\mu \nu}\epsilon_{\mu}(\gamma)\epsilon_{\nu}(\phi)\equiv t$,
 and Lorentz covariance demands $T^{\mu \nu}$ to be of the type

\begin{equation}            \label{eq:Tmunu}
T^{\mu \nu} = a \, g^{\mu \nu} + b\, P^{\mu} P^{\nu} + c\,
P^{\mu} q^{\nu}
+ d\, P^{\nu} q^{\mu} + e\, q^{\mu} q^{\nu}
\end{equation}

\noindent
where $P,\,q$ are the $\phi$ and photon momenta respectively.
Gauge invariance  $(q_{\nu}T^{\mu\nu}=0)$ forces $b=0$ and
$d=-a/(P\cdot q)$. If one works in the Coulomb gauge only the
$g^{\mu\nu}$ term of Eq.~(\ref{eq:Tmunu}) contributes which
is calculated from the $d$ term to which only the diagrams
$\textrm{b), c)}$ of Fig.~\ref{looplot} contribute. Since two powers of
momenta correspond to $P^\nu q^\mu$ the loop integrals with
the remaining momenta are actually convergent
\cite{pestieau,close}. Furthermore, if the
$K^+K^-\to\pi^0\pi^0$ amplitude is separated into the on
shell part (setting $p^2=m^2$ for the internal lines) and the
off shell remainder, it was proved in \cite{oller} that this
latter part does not contribute to the $d$ coefficient. This
information is of much value, since it allows to factorize
the on shell meson meson amplitude outside the loop integral.
The amplitude for the process is given by 

\begin{equation} \label{eq:loopsuge}
t=-\sqrt{2}\frac{e}{f^2}\epsilon(\phi)\cdot \epsilon(\gamma) \left[
M_{\phi}G_V\tilde G(M_I)+q%
\left(\frac{F_V}{2}-G_V\right)G(M_I) \right]
t_{K^+K^-,\pi^0\pi^0}
\end{equation}
where $f=92.4\textrm{ MeV}$ and 
$\tilde{G}$ is the convergent loop function of
\cite{pestieau,close} given by

\begin{eqnarray} \tilde G(M_{\phi}, M_I) =
\frac{1}{8 \pi^2} (a-b) I(a,b) \nonumber \\ a=
\frac{M_{\phi}^2}{m^2}; \quad b=
\frac{M_I^2}{m^2} 
\end{eqnarray}

\noindent
where $m$ is the mass of the meson in the loop,
$M_I$ the invariant mass of the two mesons
and $I(a,b)$ accounts for the three meson loop and is
given analytically in Eq.~(12) of \cite{oller}.
On the other hand,
$G(M_I)$ is the ordinary loop function of two meson
propagators which appears in the study of the meson meson
interaction in \cite{npa} and which is regularized there with
a cut off of the order of $1\,$GeV. In 
Eq.~(\ref{eq:loopsuge}) the  $t_{K^+K^-,\pi^0\pi^0}=
\frac{1}{\sqrt{3}}t^{I=0}_{K\overline{K},\pi\pi}$ is the
transition amplitude with the iterated loops
implicit in the coupled channels Bethe Salpeter equation (BS)
 obtained  in
\cite{npa}. 
The parameters $F_V$, $G_V$, for the vector mesons are
obtained from their decay into $e^+e^-$, $\mu^+\mu^-$ or two
mesons. If one works at the tree level one finds
$F_V^{(\rho)}=153\pm 2\textrm{ MeV}$ from $\rho\to e^+e^-$ or 
$F_V^{(\rho)}=143\pm 5\textrm{ MeV}$ from $\rho\to \mu^+\mu^-$
decay. On the other hand, from $\pi^+\pi^-$ decay at tree
level one finds $G_V^{(\rho)}=69 \textrm{ MeV}$ or if one meson
loop is considered $G_V^{(\rho)}=55 \textrm{ MeV}$ \cite{ecker}.
Similarly for the $\phi$ one finds
 $F_V^{(\phi)}=161\pm 2\textrm{ MeV}$ from $\phi\to e^+e^-$
  or  $F_V^{(\phi)}=151\pm 6\textrm{ MeV}$
  from $\phi\to \mu^+\mu^-$. From $\phi\to K^+K^-$ decay
  at tree level we find $G_V^{(\phi)}=55 \textrm{ MeV}$ but
  this value would be $5\%$ higher from $\phi\to K^0\bar{K}^0$.
It is interesting to recall that when the iteration of loops
implicit in the BS equation is used to evaluate pion and kaon
electromagnetic form factors in \cite{juan} one finds a
universal value for $G_V=55\textrm{ MeV}$ and similarly 
 $F_V=154\textrm{ MeV}$. Since here we use a similar formalism
 to the one used in \cite{juan}  the use of this latter values
 of the constants according to the approach used gives us an
 indication of the level of uncertainty in these parameters.
 We thus take for the calculations 
 $F_V=156\pm 5\textrm{ MeV}$ and  $G_V=55\pm 5\textrm{ MeV}$.

\subsection{Sequential vector meson exchange mechanisms
\label{sec:VMDtree}}

Following the lines of \cite{escribano,palomar} in the
study of $\rho$ and $\omega$ radiative decays and the more
recent of \cite{achasovlast,lucio} in the $\phi$ decay,
we also include these mechanisms here. They are depicted in
Fig.~2, where we explicitly assume that the
$\phi\to\rho^0\pi^0$ proceeds via the $\phi-\omega$ mixing.

\begin{figure}[tbp]
\centerline{\hbox{\psfig{file=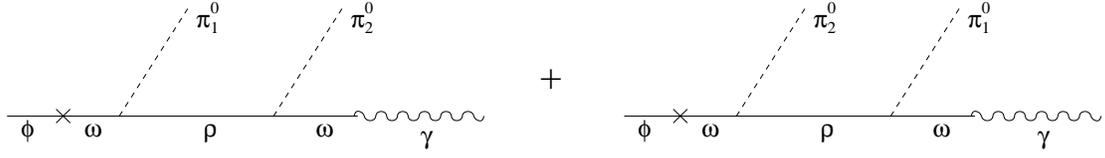,width=15cm}}}
\caption{\rm 
Diagrams for the tree level VMD mechanism.}
 \label{fig:VMD_tree}
\end{figure}

In order to evaluate these diagrams
 we use the same Lagrangians as in \cite{bramon5,escribano}

\begin{eqnarray}  \label{eq:LVVP}
{\cal L}_{VVP} &=& \frac{G}{\sqrt{2}}\epsilon^{\mu \nu \alpha \beta}\langle
\partial_{\mu} V_{\nu} \partial_{\alpha} V_{\beta} P \rangle \nonumber \\
{\cal L}_{V \gamma} &=& -4f^{2}egA_{\mu}\langle QV^{\mu}\rangle
\label{lagr}
\end{eqnarray}

\noindent
where $<>$ means $SU(3)$ trace,
$g=-4.41$,
 $G=\frac{3g^2}{4\pi^2f}=0.016\textrm{ MeV}^{-1}$,
 $Q=\textrm{diag}\{2/3,-1/3,-1/3\}$, $V$ $(P)$ the
 vector (pseudoscalar) $SU(3)$ matrices
  and $e$ is taken positive.

In addition we must use the Lagrangians producing the 
$\phi$-$\omega$ mixing. We use the formalism of \cite{urech}

\begin{equation}  
{\cal L}_{\phi\omega} = \Theta_{\phi\omega}
\,\phi_{\mu}\omega^{\mu}
\end{equation}
\noindent 
which means that the diagrams of Fig.~\ref{fig:VMD_tree}
 can be evaluated
assuming the decay of the $\omega$ (with mass $M_{\phi}$)
multiplying the amplitude
 by $\tilde{\epsilon}$ (the measure of the
$\phi$-$\omega$ mixing) given by

\begin{equation}
\tilde{\epsilon} = 
\frac{\Theta_{\phi\omega}}{M_{\phi}^2-M_{\omega}^2}
\end{equation}

Values of $\Theta_{\phi\omega}$ of the order of
$20000-29000\textrm{ MeV}^2$ are quoted in
\cite{kozhe} which are compatible with 
$\tilde{\epsilon}=0.059\pm 0.004$
used in \cite{bramonepsi}
\footnote{Note that in \cite{grau} a different sign for
$\tilde{\epsilon}$ is claimed. This is
 actually a misprint and the
results of that paper are calculated with
 $\tilde{\epsilon}>0$ \cite{bramonprivate}.}
 which is the value used here.

The amplitude for the
$\phi(q^*)\to\pi^0_1(p_1)\pi^0_2(p_2)\gamma(q)$ decay
corresponding to the diagrams of Fig.~2 is given
by

\begin{equation}
t=-{\cal C}\tilde{\epsilon}
\frac{2\sqrt{2}}{3}\frac{egf^2G^2}{M_{\omega}^2}
\left[\frac{P^2\{a\}+\{b(P)\}}{M_{\rho}^2-P^2-iM_\rho
\Gamma_\rho(P^2)}
+\frac{P'^2\{a\}+\{b(P')\}}{M_{\rho}^2-P'^2-iM_\rho
\Gamma_\rho(P'^2)}
\right]
\label{eq:VMDtree1}
\end{equation}
where $P=p_2+q$, $P'=p_1+q$ and
\begin{eqnarray} \label{eq:VMDtree2}
\{a\}&=&\epsilon^*\cdot\epsilon \ q^*\cdot q
- \epsilon^*\cdot q \ \epsilon\cdot q^* \\
\{b(P)\}&=&-\epsilon^*\cdot\epsilon \ q^*\cdot P \ q\cdot P
- \epsilon\cdot P \ \epsilon^*\cdot P \ q^*\cdot q
+ \epsilon^*\cdot q \ \epsilon\cdot P \ q^*\cdot P
+  \epsilon\cdot q^* \ \epsilon^*\cdot P \ q\cdot P
\nonumber
\end{eqnarray}

\noindent
with $\epsilon^*$ and $\epsilon$ the polarization vectors of
the $\phi$ and the photon respectively.

At this point it is worth mentioning that the theoretical
expression for the $V\to P\gamma$
decay widths $\Gamma_{V \to P\gamma}=\frac{4}{3} \alpha
 C_i^2 \left(\frac{G g f^2}{M_\rho M_V}\right)^2k^3$, 
with $C_i$ the $SU(3)$ coefficients given 
in Table I of \cite{roca}
 obtained from the Lagrangians
of Eq.~(\ref{eq:LVVP}), gives slightly different results
to the experimental values from \cite{pdg}.
 For this reason we can follow a similar procedure to that 
used for the $\eta\to\pi^0\gamma\gamma$ decay in \cite{roca}
where the $C_i$ coefficients were normalized
so that the theoretical $V\to P\gamma$ decay widths agree with
experiment. In the $\phi\to\pi^0\pi^0\gamma$ reaction
this procedure results in including in   
Eq.~(\ref{eq:VMDtree1}) a normalizing
factor ${\cal C}=0.869\pm 0.014$, obtained considering the
$V\to P\gamma$ reactions shown in Table I of \cite{roca}.

\subsection{Pion final 
state interaction in the sequential vector
meson mechanism \label{sec:VMDpiloops}}

Since the $\pi\pi$ interaction is strong in the region of
invariant masses relevant in the present reaction we next
consider the final state interaction of the pions in the
sequential vector meson mechanism. 

\begin{figure}[tbp]
\centerline{\hbox{\psfig{file=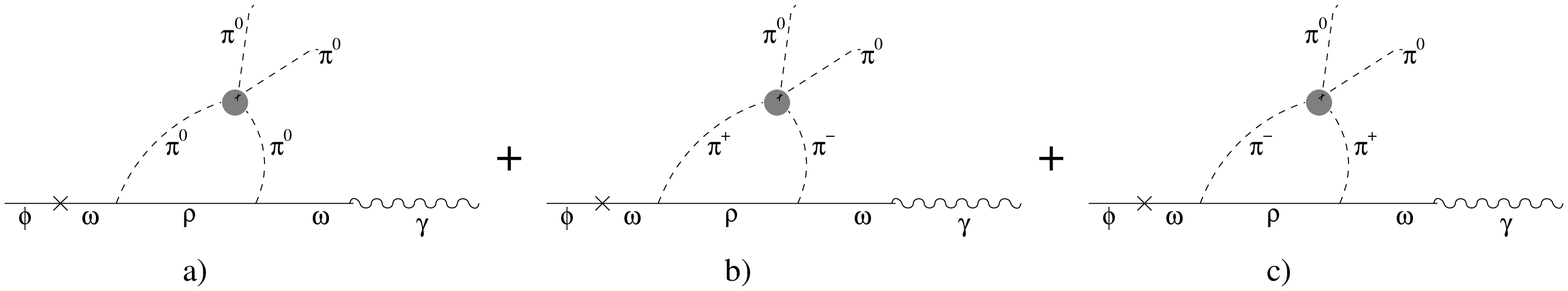,width=16cm}}}
\caption{\rm 
VMD diagrams with final state interaction of pions}
\label{fig:VMD_loop_pions}
\end{figure}

 We must take into account the loop function of
Fig.~\ref{fig:VMD_loop_pions}a,
 but on the same footing we must also consider
those of Fig.~\ref{fig:VMD_loop_pions}b
and \ref{fig:VMD_loop_pions}c, where charged pions are
produced and allowed to interact to produce the $\pi^0\pi^0$
final state. The thick dot in Fig.~\ref{fig:VMD_loop_pions}
means that one is considering the full $\pi\pi\to\pi\pi$
t-matrix, involving the loop resummation of the BS equation of
ref.~\cite{npa} and not just the lowest order 
$\pi\pi\to\pi\pi$ amplitude.

In order to evaluate those diagrams we must calculate the
loop function with a $\rho$ and two pion propagators. First
let us note that due to isospin symmetry the
$\omega\rho^0\pi^0$ coupling is the same as the
$\omega\rho^+\pi^-$ or  $\omega\rho^-\pi^+$. Next, given the
structure of the terms in Eqs.~(\ref{eq:VMDtree1}) and
(\ref{eq:VMDtree2}) we must evaluate the loop integrals 

\begin{equation}   \label{eq:loop3mesons}
i\int\frac{d^4P}{(2\pi)^4} \,P^{\mu}P^{\nu}
\frac{1}{P^2-M_V^2+i\epsilon}
\,\frac{1}{(q^*-P)^2-m_1^2+i\epsilon}
\,\frac{1}{(q-P)^2-m_2^2+i\epsilon}
\end{equation}

For simplicity we evaluate these integrals in the
reference frame where  the two meson system has zero momentum. In
this frame the $\phi$ and photon trimomentum are the same,
$\vec{q}$. We need the following integrals, which for
dimensional reasons we write as

\begin{eqnarray} \label{eq:Is}
i\int\frac{d^4P}{(2\pi)^4} P^0P^0 D_1 D_2 D_3 &=& I_0\\
\nonumber
i\int\frac{d^4P}{(2\pi)^4} P^0P^i D_1 D_2 D_3
 &=& \frac{q^i}{|\vec{q}|}I_1\\
\nonumber
i\int\frac{d^4P}{(2\pi)^4} P^iP^j D_1 D_2 D_3 
&=& \delta_{ij}I_a+\frac{q^iq^j}{\vec{q}\,^2}I_b
\end{eqnarray}
where $D_1$, $D_2$, $D_3$ are the three meson propagators of 
Eq.~(\ref{eq:loop3mesons}).
 The integrals are evaluated in the
Appendix. For that purpose the $P^0$ integral is evaluated
analytically and the $d^3\vec{P}$ integral is evaluated
numerically by means of the same cut off which has been used
to regularize the two meson loop in the meson meson
interaction, i.e., a cut off of the order of $1\,$GeV.

The amplitude is then written as 

\begin{equation}  \label{eq:tloops}
t=-{\cal C}\tilde{\epsilon}\frac{2\sqrt{2}}{3}
\frac{egf^2G^2}{M_\omega^2} \epsilon^*\cdot\epsilon
q\left\{ I_a(2q^{*0}-q) + I_bq^{*0}
+I_0q-I_1(q^{*0}+q)\right\}
2t_{\pi\pi,\pi\pi}^{I=0}
\end{equation}
where
\begin{equation}
q=\frac{M_{\phi}^2-M_I^2}{2M_I}\quad ,
\qquad q^{*0}=M_I+q=\frac{M_{\phi}^2+M_I^2}{2M_I}.
\end{equation}

Note that, since we evaluate the amplitudes in the Coulomb
gauge and $\vec{\epsilon}$ is transverse to $\vec{q}$,
$\vec{\epsilon}$ is not modified by a boost connecting 
the two meson system and
the $\phi$ rest frame. The same thing happens
to $\vec{\epsilon}\,^*$

 Thus the invariant $t$ matrix of
 Eq.~(\ref{eq:tloops})
is evaluated with the polarizations in the $\phi$ rest frame,
like the amplitude in sections \ref{sec:chiral_loops}
and \ref{sec:VMDtree}.
 The function multiplying
$\epsilon^*\cdot\epsilon$, which is thus invariant, is
evaluated in the rest frame of the two mesons, as we have
done.

In Eq.~(\ref{eq:tloops}) we have also taken into account that

\begin{equation}  
<\pi^0\pi^0+\pi^+\pi^-+\pi^-\pi^+|t|\pi^0\pi^0>
= 2 t_{\pi\pi,\pi\pi}^{I=0}
\end{equation}
where $t_{\pi\pi}^{I=0}$ is the $I=0$ $\pi\pi$ scattering
amplitude in the unitary normalization of the states used in
\cite{npa}. Note that we have used the prescription in which
the $\pi\pi$ scattering amplitude is factorized on shell in
the loops and this is justified in \cite{npa} for the loops
implicit in the BS equation, in \cite{oller} for the loops of 
Fig.~\ref{looplot}, as we mentioned before, and in
 \cite{ollernpa629} for looping diagrams of the type of those
we are now discussing. Note also that we take only the s-wave
part of the $\pi\pi\to\pi\pi$ scattering amplitude, the p-wave
part for $\pi^0\pi^0$ being forbidden. Anticipating results,
we should mention that in the case of $\pi^0\eta$ in the final
state we shall also use only the s-wave scattering amplitude
since  the p-wave part is zero at lowest order $\chi PT$ and
negligible if higher orders are considered \cite{meissner}.

\subsection{Kaon loops in the sequential vector
 meson mechanisms \label{sec:VMDKloops}}

Next we consider the diagrams analogous to those in Fig.~3 but
with kaons in the intermediate states.

\begin{figure}[tbp]
\centerline{\hbox{\psfig{file=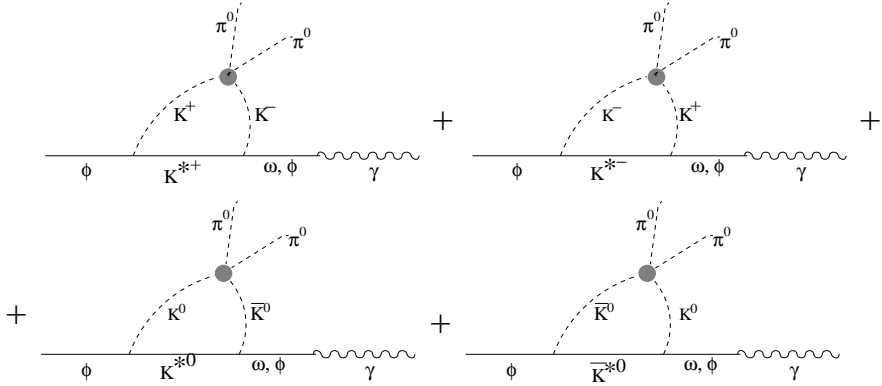,width=12cm}}}
\caption{\rm 
VMD diagrams with final state interaction of kaons}
\label{fig:VMD_loop_kaons}
\end{figure}

Note that unlike in sections \ref{sec:VMDtree}
 and  \ref{sec:VMDpiloops} the $\phi VP$ vertices
are now not OZI forbidden. They come from the Lagrangian of 
Eq.~(\ref{eq:LVVP}).
All the four $\phi K K^*$ vertices in Fig.~4 have the same
strength and taking into account the $\omega\gamma$,
$\phi\gamma$ couplings coming from Eq.~(\ref{eq:LVVP})
 we obtain
for the set of diagrams of Fig.~\ref{fig:VMD_loop_kaons}

\begin{eqnarray}  \label{eq:tKloops}
t&=&-{\cal C}
\frac{egf^2G^2}{3 } \epsilon^*\cdot\epsilon
q\left\{ I_a (2 q^{*0}-q) + I_bq^{*0}
+I_0q-I_1(q^{*0}+q)\right\}\cdot \\
\nonumber
&\cdot&\frac{4}{\sqrt{3}} t_{K\overline{K},\pi\pi}^{I=0}
\left( \frac{1}{M_\omega^2}-\frac{2}{M_{\phi}^2}\right)
\end{eqnarray}
\noindent
where the $I_i$ integrals are now evaluated using $K$,
$\bar{K}$ and $K^*$ in the three meson loop. 
In Eq.~\ref{eq:tKloops} we have used that

\begin{equation}   
< K^+K^- + K^-K^+ + K^0\overline{K^0} + \overline{K^0}K^0
|t|\pi^0\pi^0>
= \frac{4}{\sqrt{3}} t_{K\overline{K},\pi\pi}^{I=0}
\end{equation}
with $t_{K\overline{K},\pi\pi}^{I=0}$ the $I=0$
$K\overline{K}\to\pi\pi$ 
unitarized transition amplitude evaluated with
the states with the unitary normalization of \cite{npa}.
Unlike in  the case of pion loops where only the $\omega$ is
attached to the photon, now we can have the $\omega$ and the
$\phi$ without violating the OZI rule. The two terms in the
last factor of Eq.~(\ref{eq:tKloops}) account
for the $\omega$
and $\phi$ meson respectively.

\subsection{Sequential axial vector meson mechanisms}

Since the mass of the $\phi$ is around $250\textrm{ MeV}$
higher than the $\rho$ mass, and we are considering sequential
vector meson mechanisms with $\rho$ or $K^*$ exchange, we
should pay attention to the analogous mechanisms involving
vector mesons with a similar mass difference with the $\phi$
on the upper side and these are the axial and vector mesons
with $J^{PC}=1^{+-}$ or $1^{++}$ (see
Table~1). Therefore,  the $b_1$ or $a_1$ axial vector
 mesons and the
$K_{1B}$, $K_{1A}$ strange axial vector mesons will play the
role of the $\rho$ or the $K^*$ in former diagrams.

\begin{table}[htbp]
\begin{center}
\begin{tabular}{|c||c||c|c|}\hline 
$J^{PC}$ &$I=1$  &$I=0$ & $I=1/2$ \\ \hline \hline  
 $1^{+-}$  & $b_1(1235)$  & $h_1(1170)$, $h_1(1380)$
    &$K_{1B}$ \\ \hline  
 $1^{++}$  & $a_1(1260)$  & $f_1(1285)$, $f_1(1420)$
    &$K_{1A}$ \\ \hline     
\end{tabular}
\end{center}
\caption{Octets of axial-vector mesons.}
\end{table}

Because of the $C$
parity of the states, the Lagrangians for
the axial-vector--vector--pseudoscalar couplings have
the structure of $<B\{V,P\}>$ for the $b_1$ octet and 
$<A[V,P]>$ for the octet of the $a_1$ \cite{gatto}, where
the $<>$ means $SU(3)$ trace. In the last expressions $V$ and
$P$ are the usual vector and pseudoscalar $SU(3)$ matrices
respectively and 

\begin{eqnarray}
B &\equiv& \left(\begin{array}{ccc} 
\frac{1}{\sqrt{2}}h_1(1170) + \frac{1}{\sqrt{2}} b^0_1
 & b^+_1 & K_{1B}^+\\
b_1^-&\frac{1}{\sqrt{2}}h_1(1170) - \frac{1}{\sqrt{2}} b^0_1
& K^0_{1B}\\
K^-_{1B}& \overline{K}^0_{1B} & h_1(1380)
\end{array}
\right)  \nonumber \\
A &\equiv& \left(\begin{array}{ccc} 
\frac{1}{\sqrt{2}}f_1(1285) + \frac{1}{\sqrt{2}} a^0_1
 & a^+_1 & K_{1A}^+\\
a_1^-&\frac{1}{\sqrt{2}}f_1(1285) - \frac{1}{\sqrt{2}} a^0_1
& K^0_{1A}\\
K^-_{1A}& \overline{K}^0_{1A} & f_1(1420)
\end{array}
\right) . 
\end{eqnarray}

 In addition one has to consider 
an approximate $50\,\%$ mixture of the
$K_{1B}$ and $K_{1A}$ states to give the physical $K_1(1270)$
and $K_1(1400)$  states \cite{suzuki,carnegie,gatto}.

We have modified the
original Lagrangian of \cite{gatto} to treat the
vector fields in the tensor formalism of \cite{ecker}. This
formalism has the advantage that without basically changing
the rates of the $A\to VP$ decays, one deduces the coupling of
the $a_1$ to $\pi\gamma$ using vector meson dominance through
$a_1\to\pi\rho\to\pi\gamma$, with an amplitude which is gauge
invariant  and which is in agreement with the chiral structure
of \cite{ecker} for  the $a_1\to P\gamma$ couplings and with
the experiment. Details are given elsewhere in \cite{axial}.

We hence use the Lagrangians \cite{axial}

\begin{eqnarray}  \label{eq:LBLA}
{\cal L}_{BVP}&=&\tilde{D} <B_{\mu\nu}\{V^{\mu\nu},P\}> \\
\nonumber
{\cal L}_{AVP}&=&i\tilde{F} <A_{\mu\nu}[V^{\mu\nu},P]> 
\end{eqnarray} 
where the $i$ factor in front of the $\tilde{F}$ is needed
in order ${\cal L}_{AVP}$ to be hermitian.

In Eq.~(\ref{eq:LBLA}) the fields 
$W_{\mu\nu}\equiv A_{\mu\nu}$, $B_{\mu\nu}$ are normalized
such that 
\begin{equation}
<0|W_{\mu\nu}|W;P,\epsilon>=
\frac{i}{M_W}\left[ P_\mu\,\epsilon_\nu(W)-P_\nu\, 
\epsilon_\mu(W)\right]
\end{equation}

In addition the propagators with the tensor fields are defined
as \cite{ecker}

\begin{eqnarray} \label{eq:prop_4indices}
&<&0|T\{W_{\mu\nu}W_{\rho\sigma}\}|0>=i{\cal
D}_{\mu\nu\rho\sigma} =\\ \nonumber
&=&i\frac{M_W^{-2}}{M_W^2-P^2-i\epsilon}\left[g_{\mu\rho}\,
g_{\nu\sigma}\,(M_W^2-P^2)
+g_{\mu\rho}\,
P_\nu \, P_\sigma-g_{\mu\sigma}\,P_\nu \, P_\rho-(\mu \leftrightarrow \nu) \right]
\end{eqnarray}

The physical $K_1(1270)$ and $K_1(1400)$, with a mixture
around $45$ degrees
\footnote{It is worth mentioning that in \cite{suzuki,axial} two
more possible solutions for the mixing angle around $30$ and
$60$ degrees were found. This uncertainty will be taken into
account in the evaluation of the error band in our final
results.}
 found in
\cite{suzuki,carnegie,gatto,axial}, can be expressed, in
terms of the $I=1/2$ members of the $1^{+-}(1^{++})$ octets, 
$K_{1B}(K_{1A})$, as

\begin{eqnarray}
\nonumber
K_1(1270)=\frac{1}{\sqrt{2}}(K_{1B}-iK_{1A}) \\
K_1(1400)=\frac{1}{\sqrt{2}}(K_{1B}+iK_{1A})
\end{eqnarray} 

With the values for  $\tilde{D}=-1000\pm 120 \ MeV$ and 
$\tilde{F}=1550\pm 150 \ MeV$, very similar to those found in
\cite{suzuki,carnegie,gatto}, we are able to describe all the
$A\to VP$ decays plus the radiative decays of the
$a_1\to\pi\gamma$ \cite{axial}.

Once again the $\phi$ sequential decay
at tree level through $b_1$ exchange
is OZI violating and proceeds via $\phi-\omega$ mixing.
We have evaluated this contribution and found it negligible,
thus we do not further discuss it.
Similarly the loops involving
pions are equally negligible.
 On the other hand $a_1$ exchange
is not allowed by $C$ and $G$ parity.

\subsection{Kaon loops from sequential axial vector meson mechanisms}

\begin{figure}[tbp]
\centerline{\hbox{\psfig{file=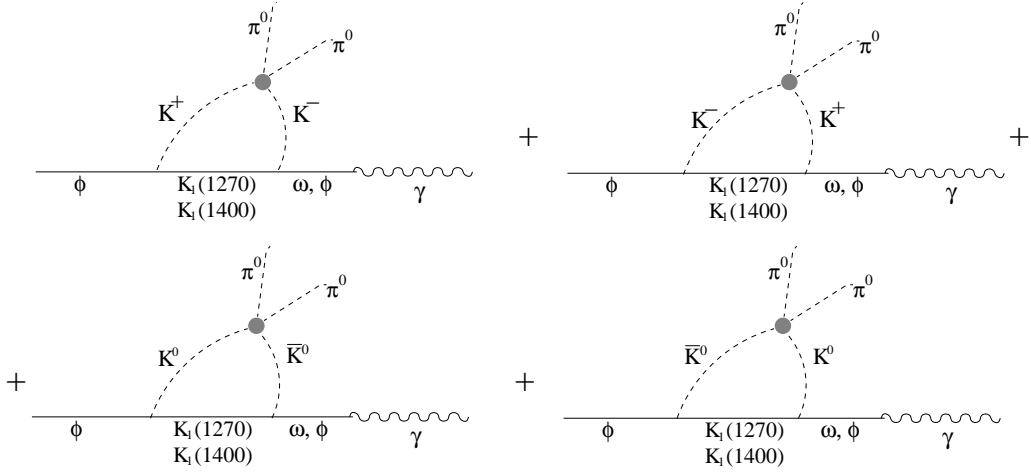,width=14cm}}}
\caption{\rm 
Diagrams for the sequential mechanisms involving the $K_1$
axial-vector mesons.}
\label{fig:VMD_loop_K1}
\end{figure}

The relevant mechanisms involving axial-vectors
are those in Fig.~\ref{fig:VMD_loop_K1} in which
 $K$, $\bar{K}$ are created and
through scattering lead to the final $\pi^{0}\pi^{0}$ state. These are not OZI
forbidden and have a nonnegligible contribution.

Since we are using the tensor formulation for the vector
mesons this forces us to use the tensor coupling of the photon
to the vector mesons obtained from the $F_V$ term of
Eq.~(\ref{eq:LFVGV}):

\begin{equation}
{\cal L}_{V\gamma}=-e\frac{F_V}{2}\lambda_V V_{\mu\nu}^0
(\partial^\mu A^\nu - \partial^\nu A^\mu)
\end{equation}

with $\lambda_V=1,\frac{1}{3},-\frac{\sqrt{2}}{3}$ for 
$V=\rho,\omega,\phi$ respectively.

Now the amplitude with $K_1(1270)$ as intermediate state
 is given, following the lines of sections
  \ref{sec:VMDpiloops} and \ref{sec:VMDKloops} as

\begin{center}
\begin{eqnarray}\label{eq:tk1}
t&=&\frac{4eF_V}{M_\phi M_{K_1}^2}
\epsilon^*\cdot\epsilon
\frac{4}{\sqrt{3}}  t_{K\bar{K},\pi\pi}^{I=0}
 \left[\frac{(\tilde D^2-\tilde F^2)}{6\sqrt{2}M_\omega^2}
-\frac{\sqrt{2}(\tilde D+\tilde F)^2}{6M_\phi^2}\right] 
\cdot \\
\nonumber
&\cdot&
q\left[ q^{*0}I_{0}+I_{a}(2 q-q^{*0}) 
+ I_{b}q-I_{1}(q+q^{*0})+(q
-q^{*0})
G_{K\bar{K}}(M_I)\right]_{K_{1}\equiv K_{1}(1270)}
\end{eqnarray}
\end{center}

The amplitude for the diagram with intermediate $K_1(1400)$ is
the same as Eq.~(\ref{eq:tk1}) but replacing $\tilde{F}$ by
$-\tilde{F}$ and using the $K_1(1400)$ propagator in the
evaluation of the $I_i$ integrals.

The term proportional to $G_{K\bar{K}}(M_I)$ at the end of the
former expression comes from the $M_{K_1}^{2}-P^{2}$
of Eq.~(\ref{eq:prop_4indices})
 which cancels the intermediate vector
meson propagator, thus leaving
 the loop with just two pseudoscalar propagators, which is
  the same one that appears in the meson meson scattering.

\section{The $\phi\rightarrow\pi^{0} \eta \gamma$ decay}

 After the discussion of the former points the consideration of the $\phi \rightarrow \pi^{0}\eta \gamma$ decay requires only minimal technical details which we write below.

\subsection{Kaon loops from $\phi \rightarrow K \bar{K}$ decay.}

The diagrams to be considered are exactly the same as those
in Fig.~\ref{looplot} changing the final $\pi^{0} \pi^{0}$ by
$\pi^{0}\eta$. Technically all that we do is to substitute 
in Eq.~(\ref{eq:loopsuge}) the amplitude
$t_{K^+K^-,\pi^0\pi^0}$ by
the $t_{K^+K^-,\pi^0\eta}$ or equivalently
 $-\frac{1}{\sqrt{2}}t_{K\bar
K,\pi^0\eta}^{I=1}$ in the nomenclature of \cite{npa}.

\subsection{Sequential vector meson contribution}

We have now the diagrams

\begin{figure}[tbp]
\centerline{\hbox{\psfig{file=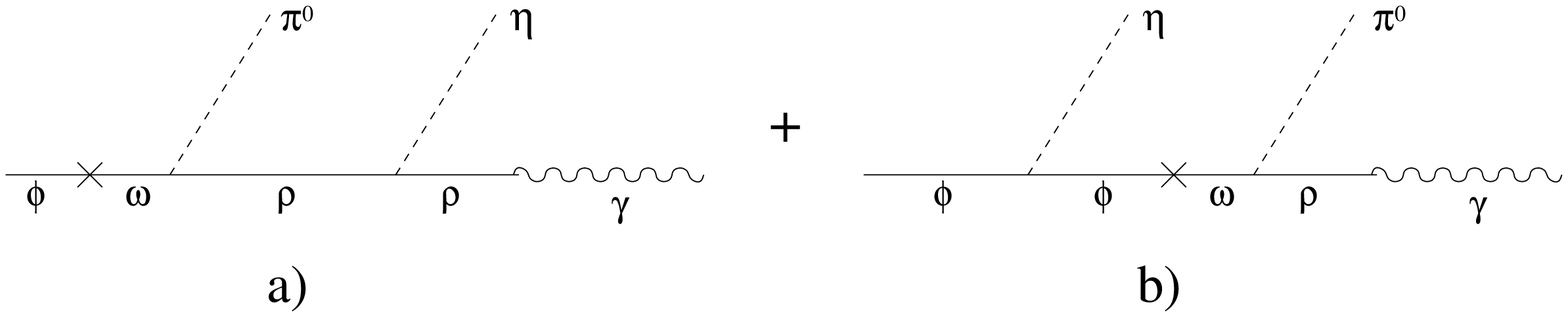,width=16cm}}}
\caption{\rm 
Diagrams for the tree level VMD mechanism for the
$\phi\to\pi^0\eta\gamma$ decay.}
\label{fig:VMD_tree_pieta}
\end{figure}

The amplitude of this contribution
is very similar to Eq.~(\ref{eq:VMDtree1}). We find

\begin{equation}
t=-{\cal C}\tilde{\epsilon}
\frac{4}{\sqrt{3}}\frac{egf^2G^2}{M_{\rho}^2}
\left\{\frac{P^2\{a\}+\{b(P)\}}{M_{\rho}^2-P^2-iM_\rho
\Gamma_\rho(P^2)}
-\frac{P'^2\{a\}+\{b(P')\}}{M_{\phi}^2-P'^2-iM_\phi
\Gamma_\phi(P'^2)}
\right\}
\label{eq:VMDtreepieta}
\end{equation}

\subsection{$\pi\eta$ final state interaction of the
sequential vector meson mechanism}

The results for the amplitude corresponding to FSI in diagram
\ref{fig:VMD_tree_pieta}a is easily obtained from 
Eq.~(\ref{eq:tloops}). We multiply the expression by 
$\sqrt{6}$, change the masses and widths of the corresponding
intermediate particles in the loops and change 
$2t_{\pi\pi,\pi\pi}^{I=0}$ by
$t_{\pi\eta,\pi\eta}^{I=1}$. The
 amplitude corresponding to the FSI in diagram
\ref{fig:VMD_tree_pieta}b is done in the same way but we
multiply by $-\sqrt{6}$ instead 
of $\sqrt{6}$ to account for the
negative sign in Eq.~(\ref{eq:VMDtreepieta}).

\subsection{Kaon loops from sequential vector meson
mechanisms }

The diagrams involved are the same as in 
Fig.~\ref{fig:VMD_loop_kaons} but instead of $\omega$ and
 $\phi$ coupled to the photon we have now only the $\rho$ and
 the final $\pi^0\pi^0$ is substituted by $\pi^0\eta$.
 Altogether one must change $\frac{1}{M_\omega^2}$
 to $\frac{1}{M_\rho^2}$ in the $\omega$ term of
  Eq.~(\ref{eq:tKloops}) and substitute 
  $t_{K\bar K,\pi\pi}^{I=0}$ by 
  $-3\sqrt{\frac{3}{2}}t_{K \bar K,\pi\eta}^{I=1}$.

\subsection{Axial vector meson exchange mechanisms}

The tree level with $b_1$, $a_1$ exchange and their octet
partners are also OZI forbidden and negligible 
like in the case
of $\pi^0\pi^0$ final state. However, the term with kaon loops
equivalent to Fig.~\ref{fig:VMD_loop_K1} but with $\pi^0\eta$
in the final state and a $\rho$ coupled to the photon instead
of $\omega$, $\phi$ are OZI allowed and we take them into
account.

Once again the amplitude is easily obtained from that of the
$\omega$ part of Eq.~(\ref{eq:tk1}) by changing 
$\frac{1}{M_\omega^2}$ to $\frac{1}{M_\rho^2}$ and 
 $t_{K\bar K,\pi\pi}^{I=0}$ to 
 $-\frac{3\sqrt{3}}{\sqrt{2}}t_{K \bar K,\pi\eta}^{I=1}$ and
 the same considerations  regarding the contribution of the 
 $K_1(1270)$ and $K_1(1400)$ intermediate states.

\section{Results}

\subsection{Differential cross section}
Using the transition amplitudes described in the previous sections, we
can calculate the differential decay widths of the $\phi$
meson as,

\begin{equation}
\frac{d\Gamma}{dM_I}= \frac{1}{64 \pi^3}
 \int_{m_\pi}^{M_\phi-q-m^\prime}
d\omega \frac{M_I}{M_\phi^2} \bar{\sum} \sum | t |^2
 \Theta(1-\cos^2\theta_{\pi^0\gamma}),
\label{dismi}
\end{equation}

\noindent
where $M_I$ is the invariant mass of the final two mesons,
$m^\prime$ is the
pion mass for the $\pi\pi\gamma$ decay and 
the $\eta$ mass for the
$\pi\eta\gamma$ decay and $q$ is the photon momentum in the
initial $\phi$ rest frame. $\Theta$ is the step function
and $\theta_{\pi^0\gamma}$ accounts for the
angle between 
the $\pi^0$ and the photon and it is given by

\begin{equation}
\cos^2\theta_{\pi^0\gamma}  =\frac{1}{2pq} \left[ (M_\phi-\omega(p)-q)^2 - m^{\prime 2} -
p^2 - q^2 \right],
\end{equation}

\noindent
where $p$ and $\omega(p)$ are the $\pi^0$ momentum and energy
 in the initial $\phi$ rest frame. A
symmetry factor 1/2 must be implemented in 
Eq.~(\ref{dismi}) in the case of
$\pi^{0} \pi^{0}$ in the final state.

The spin sum and average of the transition amplitudes, $\bar{\sum} \sum
| t |^2$, can be expressed using the contravariant tensor $F^{ij}$ as:

\begin{equation}
\bar{\sum} \sum | t |^2 = \frac{1}{3} \left[
F^{ij} F^{ij*} - \frac{1}{|q|^2} (F^{ij}q_j) (F^{i j^\prime *}
q_{j^\prime})
\right],
\end{equation}

\noindent
where the tensor expression $F^{ij}$ of the transition amplitude $t$ is
defined as

\begin{equation}
t \equiv F^{ij} \epsilon_i (V) \epsilon_j (\gamma).
\end{equation}

\subsection{Results for the $\phi\to\pi^0\pi^0\gamma$ decay}

\begin{figure}[tbp]
\centerline{\hbox{\psfig{file=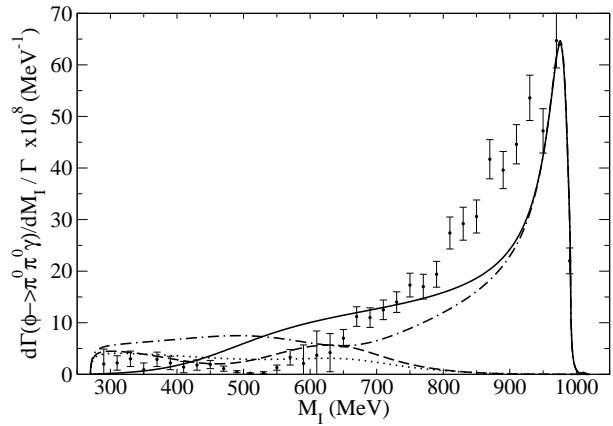,width=9cm,angle=-90}}}
\caption{\rm 
Several contributions to the $\pi^0\pi^0$ invariant mass distribution
for the $\phi\to\pi^0\pi^0\gamma$ decay.
Solid line: chiral loops of Fig.~\ref{looplot}.
Dotted line: sequential VMD at tree level of Fig.~\ref{fig:VMD_tree}.
Dashed line: sequential VMD including the FSI of Fig.~\ref{fig:VMD_loop_pions}.
Dashed-dotted line: Coherent sum of all these mechanisms.
(Experimental data from \cite{frascati1}).
 }
\label{fig:res1}
\end{figure}

First of all, we show in Fig.~\ref{fig:res1} the contribution
of the chiral loops (solid line) together with the sequential
VMD mechanisms at tree level of Fig.~\ref{fig:VMD_tree}
(dotted line) and the VMD with final state interaction,
Fig.~\ref{fig:VMD_loop_pions}, (dashed line). In the
dashed-dotted line we have plotted the contribution of all
these mechanisms together. We can see, that the effect of the
VMD tree level contribution and its unitarization is small
compared to the chiral loops from  $\phi\to K^+K^-$ decay.
The total contribution is still rather different than the
experimental values of \cite{frascati1}, although the inclusion of
the sequential vector meson mechanism provides some strength
for the distributions at low invariant masses where the kaon
loops from $\phi\to K^+K^-$ are negligible.

\begin{figure}[tbp]
\centerline{\hbox{\psfig{file=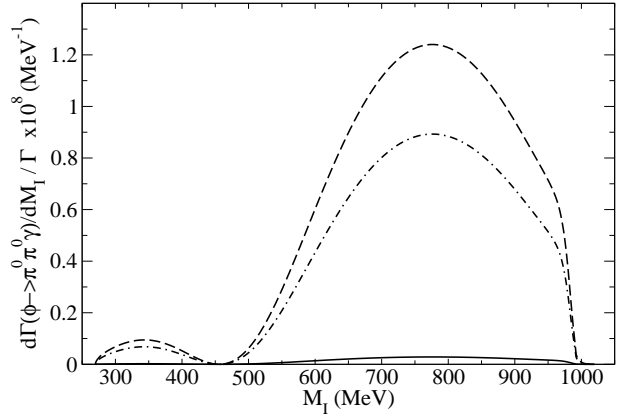,width=9cm,angle=-90}}}
\caption{\rm 
Contribution of the sequential mechanisms involving kaon
loops. Dashed line: $\phi$ contribution. Dashed-dotted line:
$\omega$ contribution. Solid line: Coherent sum of $\phi$ and
$\omega$ contributions.
}
\label{fig:res2}
\end{figure}

\begin{figure}[tbp]
\centerline{\hbox{\psfig{file=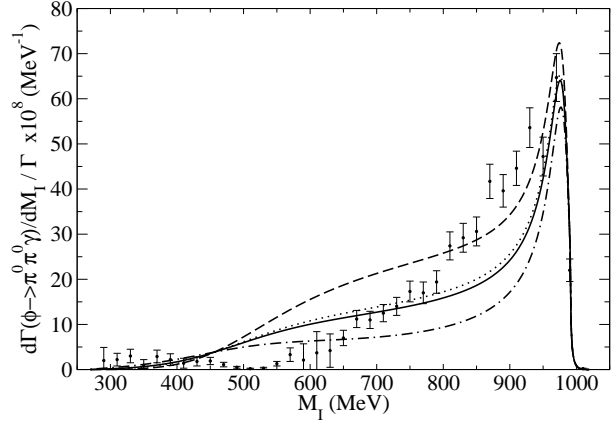,width=9cm,angle=-90}}}
\caption{\rm 
Addition of the sequential loops contribution to the 
chiral loops of Fig.~\ref{looplot}.
Solid line: chiral loops of Fig.~\ref{looplot}.
Dashed line: chiral loops of Fig.~\ref{looplot} + sequential
loops involving kaons with $\phi$. 
Dashed-dotted line: chiral loops of Fig.~\ref{looplot} + sequential
loops involving kaons with $\omega$.
Dotted line: Coherent sum of all these contributions.
}
\label{fig:res3}
\end{figure}

In Fig.~\ref{fig:res2} we can see the contribution of the
sequential mechanisms involving kaon loops
(Fig.~\ref{fig:VMD_loop_kaons}), separating the $\phi$
(dashed line) and $\omega$ (dashed-dotted line)
contributions, and the coherent sum of both amplitudes (solid
line). We can see that the sum of both mechanisms almost
cancels,  thus making the contribution of the kaon loops of
the sequential VMD mechanisms very small, although the
individual contribution of the loops with an $\omega$ or a
$\phi$ meson attached to the photon are sizeable when they
interfere separately with the chiral loops from $\phi\to
K^+K^-$ decay, as it can be seen in Fig.~\ref{fig:res3}. Had
this accidental cancellation  not happened the contribution
of the kaon loops of the sequential mechanisms would have
been sizeable.

\begin{figure}[tbp]
\centerline{\hbox{\psfig{file=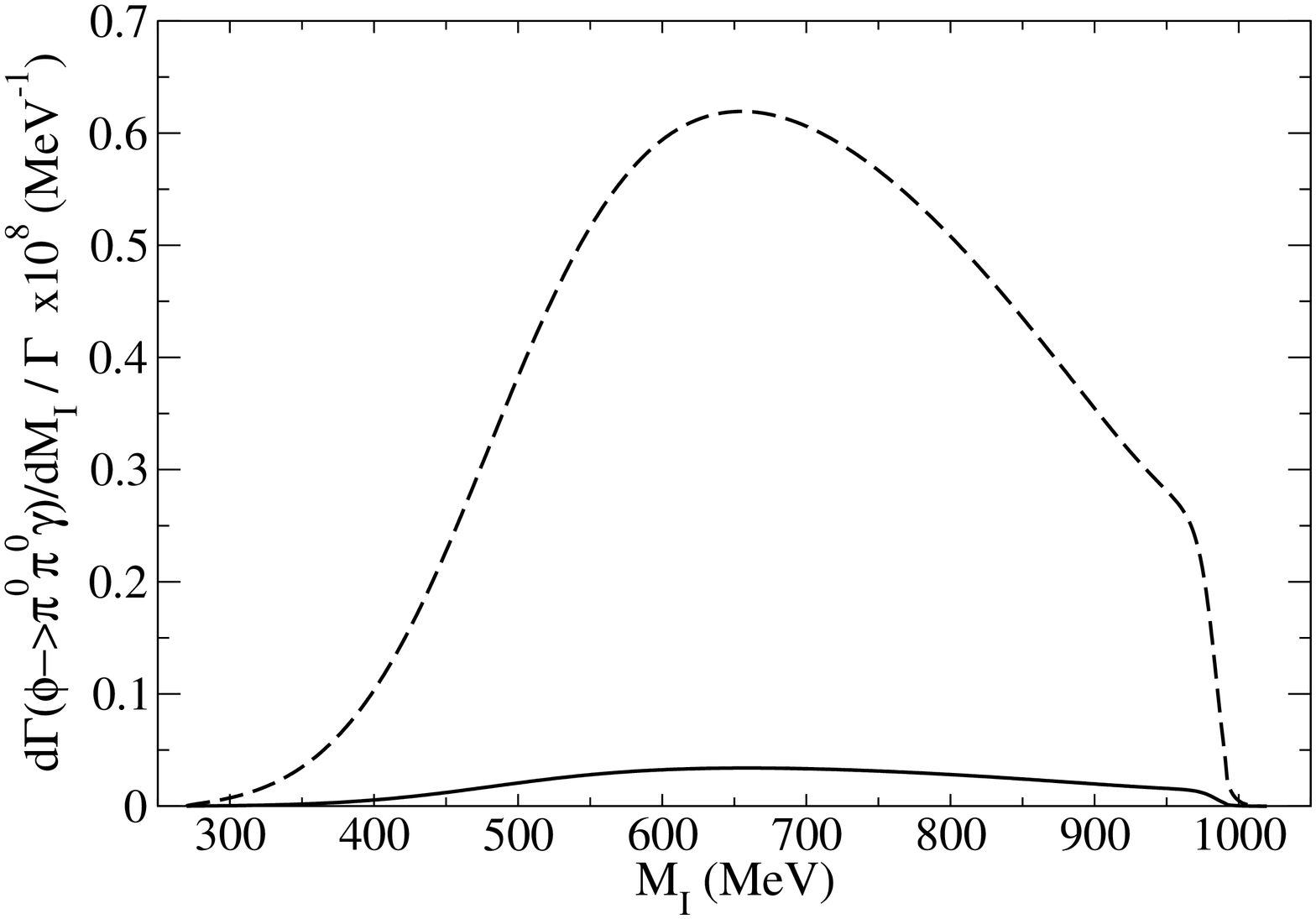,width=9cm,angle=-90}}}
\caption{\rm 
Contribution of the sequential mechanisms involving
axial-vector mesons.
Solid line: $K_1(1270)$ contribution.
Dashed line: $K_1(1400)$ contribution.
}
\label{fig:res4}
\end{figure}

\begin{figure}[tbp]
\centerline{\hbox{\psfig{file=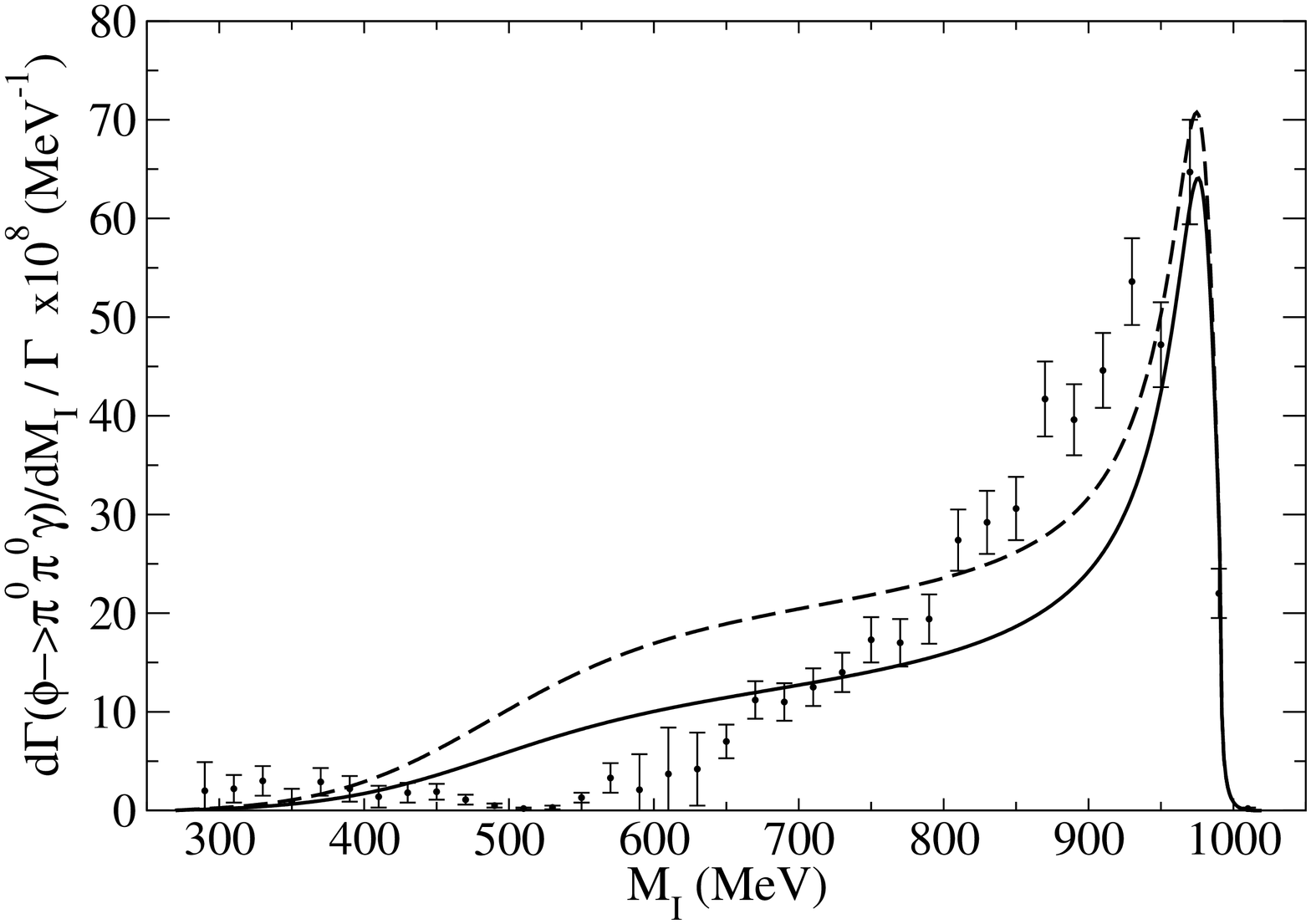,width=9cm,angle=-90}}}
\caption{\rm
 Addition of the sequential loops involving axial-vector
 mesons to the 
chiral loops of Fig.~\ref{looplot}.
Solid line: chiral loops of Fig.~\ref{looplot}.
Dashed line: chiral loops of Fig.~\ref{looplot} + axial-vector
 meson contribution.
}
\label{fig:res5}
\end{figure}

In Fig.~\ref{fig:res4} we show the contribution of the
axial-vector meson resonances.  We can see the contribution
of the mechanisms with a $K_1(1270)$ as intermediate state
(dashed line) and the one with $K_1(1400)$ (solid line) (see
Fig.~\ref{fig:VMD_loop_K1}). The size of these mechanisms by
themselves is very small but when they interfere with the
chiral loops  from $\phi\to K^+K^-$ decay they give an
important contribution, as shown in Fig.~\ref{fig:res5}.
Thus, these mechanisms with the axial-vector mesons in the
intermediate states cannot be neglected.

\begin{figure}[tbp]
\centerline{\hbox{\psfig{file=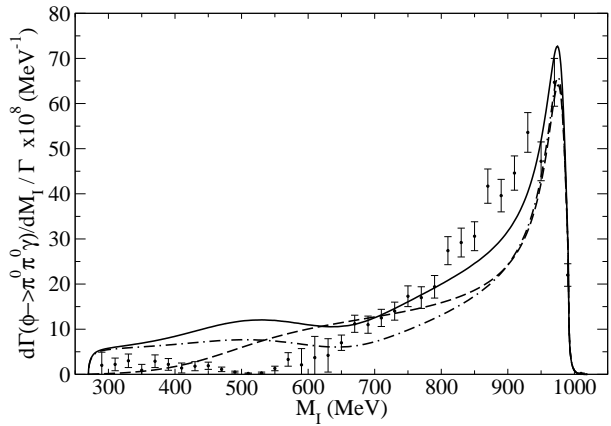,width=9cm,angle=-90}}}
\caption{\rm 
Different contributions to the two pion invariant mass
distributions of the $\phi\to\pi^0\pi^0\gamma$ decay:
Dashed line: chiral loops of Fig.~\ref{looplot}.
Dashed-dotted line: chiral loops of Fig.~\ref{looplot} + 
sequential VMD and its final
state interaction.
Solid line: idem plus the contribution of the mechanisms
involving axial-vector mesons, (full model).
}
\label{fig:res6}
\end{figure}

In Fig.~\ref{fig:res6} we show the contribution of the
different relevant mechanisms as they have been added to the
model.  In the dashed line we can see the contribution of the
chiral loops. In the dashed-dotted line we  add the
contribution of the sequential VMD and its final
state interaction. In the solid line we add the contribution
of the loops of the sequential mechanisms with axial-vector
mesons in the intermediate states. This latter line
represents the full model.

\begin{figure}[tbp]
\centerline{\hbox{\psfig{file=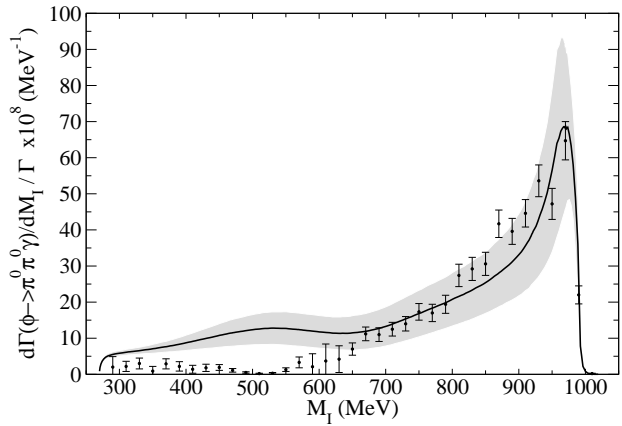,width=9cm,angle=-90}}}
\caption{\rm 
Final results for the $\pi^0\pi^0$ invariant mass distribution
for the $\phi\to\pi^0\pi^0\gamma$ decay with the theoretical
error band. Experimental data from \cite{frascati1}.
}
\label{fig:res7}
\end{figure}

Up to now, all the curves shown in the figures have been
calculated using the central values of the parameters without
considering the uncertainties in their values. 
In Fig.~\ref{fig:res7} we
show the final result but including an evaluation of the error
band due to the uncertainties in the parameters of the model.
This error band has been calculated implementing a Monte
Carlo gaussian sampling of the parameters within their
experimental errors. The parameters of the model which
uncertainties are relevant in the error analysis
are shown in Table \ref{tabla2}.

\begin{table}[tbp]
\begin{center}
\begin{tabular}{|c|c|c|c|}\hline 
${\cal C}$ &$\tilde{\epsilon}$  &$G_V$ (MeV) &
 $F_V$ (MeV) \\  
 $0.869\pm 0.014$  & $0.059\pm 0.004$  & $55\pm 5$&$156\pm 5$
   \\ \hline & & &  \\[-0.4cm]
 $f_\pi$  (MeV)& $\Lambda$ (MeV) & $\tilde{D}$ (MeV)&
  $\tilde{F}$ (MeV)
   \\
  $92.4\pm 3\%$ & $1000\pm 50$ & $-1000\pm 120$&$1550\pm150$
  \\ \hline     
\end{tabular}
\end{center}
\caption{Parameters which uncertainties are relevant in
 the error analysis. The $f_\pi$  and $\Lambda$ are the $f_\pi$
constant and cutoff of the momentum integral respectively 
in the loops involved  in the unitarized
 meson-meson rescattering.}
 \label{tabla2}
\end{table}

The errors in $f_\pi$ and $\Lambda$ assumed
 in the calculations
have been chosen such that the quality of the fit to the
$\pi\pi$ phase shifts along the lines of \cite{npa} is still
acceptable within experimental errors.

The parameter with the larger contribution to the error band
turns out to be the $G_V$ since the largest contribution,
chiral kaon loops form $\phi\to K^+K^-$ decay, is roughly
proportional to $G_V$ (up to the term with
$q(\frac{F_V}{2}-G_V)$ in Eq.~(\ref{eq:loopsuge}) which would
be zero within some vector meson dominance hypotheses
\cite{ecker} and is small with our set of parameters).

The total width and branching ratio obtained in the present
work are

\begin{eqnarray}
\Gamma_{\phi\to\pi^0\pi^0\gamma}&=&520\pm 150 \textrm{ eV}\\
\nonumber
BR(\phi\to\pi^0\pi^0\gamma)&=&(1.2\pm 0.3)\times 10^{-4}
\end{eqnarray}
to be compared with the experimental values 

\begin{eqnarray} \nonumber
BR^{exp}(\phi\to\pi^0\pi^0\gamma)
&=&(1.22\pm 0.10\pm 0.06)\times 10^{-4} 
\quad \textrm{ \cite{novo1}}\\
 \nonumber
BR^{exp}(\phi\to\pi^0\pi^0\gamma)
&=&(0.92\pm 0.08\pm 0.06)\times 10^{-4} 
\quad \textrm{ \cite{novo2}}\\
 \nonumber
BR^{exp}(\phi\to\pi^0\pi^0\gamma)
&=&(1.09\pm 0.03\pm 0.05)\times 10^{-4}
\quad \textrm{ \cite{frascati1}}
\end{eqnarray}

In Fig.~\ref{fig:res7} we can see that our results, considering the error
band, fairly agree with the experimental data except in the 
region around $500\textrm{ MeV}$.
 The reason of this discrepancy will be further discussed in
Section \ref{sectionManolo}.

\subsection{Results for the $\phi\to\pi^0\eta\gamma$ decay}

\begin{figure}[tbp]
\centerline{\hbox{\psfig{file=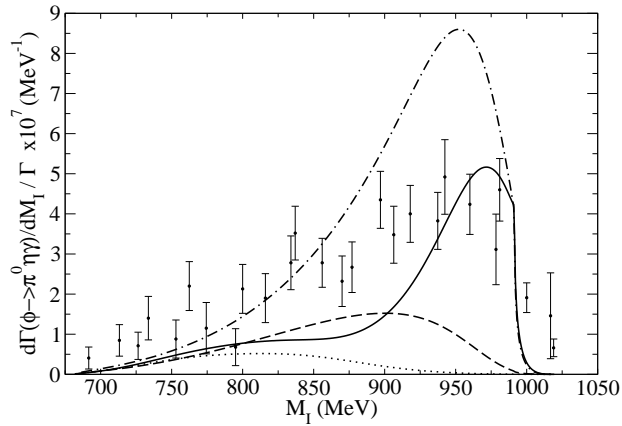,width=9cm,angle=-90}}}
\caption{\rm 
Several contributions to the $\pi^0\eta$ invariant mass distribution of
the $\phi\to\pi^0\eta\gamma$ decay.
Dashed-dotted line: chiral loops of Fig.~\ref{looplot}.
Dotted line: sequential VMD at tree level.
Dashed line: loops of the sequential VMD involving kaons.
Solid line: coherent sum of all these mechanisms.
(Experimental data from \cite{achasovlast}).}
\label{fig:res8}
\end{figure}

In Fig.~\ref{fig:res8} we show the contribution of the chiral
loops from $\phi\to K^+K^-$ decay (dashed-dotted line)
together with the sequential VMD at tree level
(dotted line) and the loops of the sequential VMD 
involving kaons (dashed line). In the solid line we show the
coherent sum of all these contributions. 

In this case the VMD mechanism involving $\pi\eta$ creation
followed by $\pi\eta\to\pi\eta$ FSI is negligible and,
 therefore,
is not shown in the figure. We can see that the kaon loops
of the sequential VMD mechanism give a very important
contribution, in contrast to the $\phi\to\pi^0\pi^0\gamma$
case where the important contribution of the VMD sequential
loops  was the one involving pions. 
Recall  that the reason for the small contribution of kaon
loops in the VMD mechanism for $\phi\to\pi^0\pi^0\gamma$ was
the accidental cancellation of the $\phi$ and
$\omega$ contributions (for the $\phi$ and $\omega$ attached
to the photon). However, in this case there is only the $\rho$
attached to the photon and thus the contribution of this term
is large, as was also the case in  $\phi\to\pi^0\pi^0\gamma$
for the individual contributions.

\begin{figure}[tbp]
\centerline{\hbox{\psfig{file=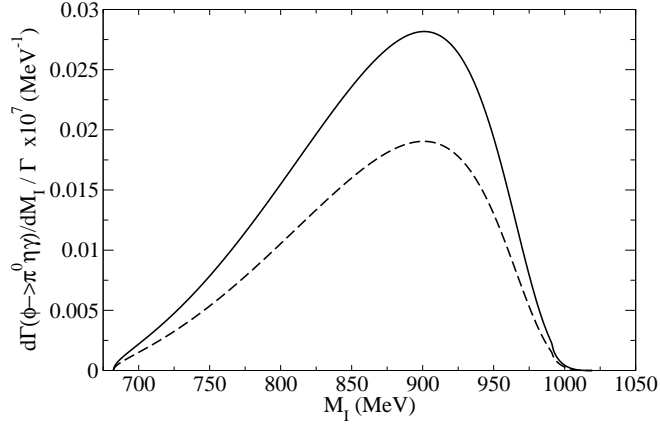,width=9cm,angle=-90}}}
\caption{\rm 
Contribution of the sequential mechanisms involving
axial-vector mesons.
Solid line: $K_1(1270)$ contribution.
Dashed line: $K_1(1400)$ contribution.
}
\label{fig:res9}
\end{figure}

In Fig.~\ref{fig:res9} we show the contribution of the
mechanisms with intermediate $K_1(1270)$ (solid line) and
$K_1(1400)$ (dashed line).

\begin{figure}[tbp]
\centerline{\hbox{\psfig{file=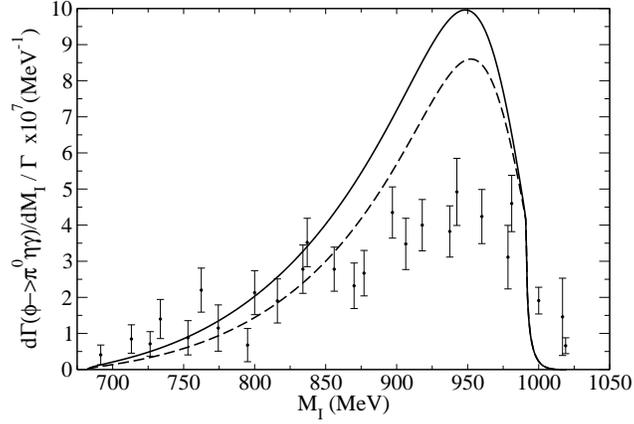,width=9cm,angle=-90}}}
\caption{\rm 
 Addition of the sequential loops involving axial-vector
 mesons to the 
chiral loops of Fig.~\ref{looplot}.
Solid line: chiral loops of Fig.~\ref{looplot}.
Dashed line: chiral loops of Fig.~\ref{looplot} + axial-vector
meson contribution.
}
\label{fig:res10}
\end{figure}

These contributions by themselves are very small but when
they interfere with the chiral loops give a sizeable
contribution, (from dashed to solid line in
Fig.~\ref{fig:res10}).

\begin{figure}[tbp]
\centerline{\hbox{\psfig{file=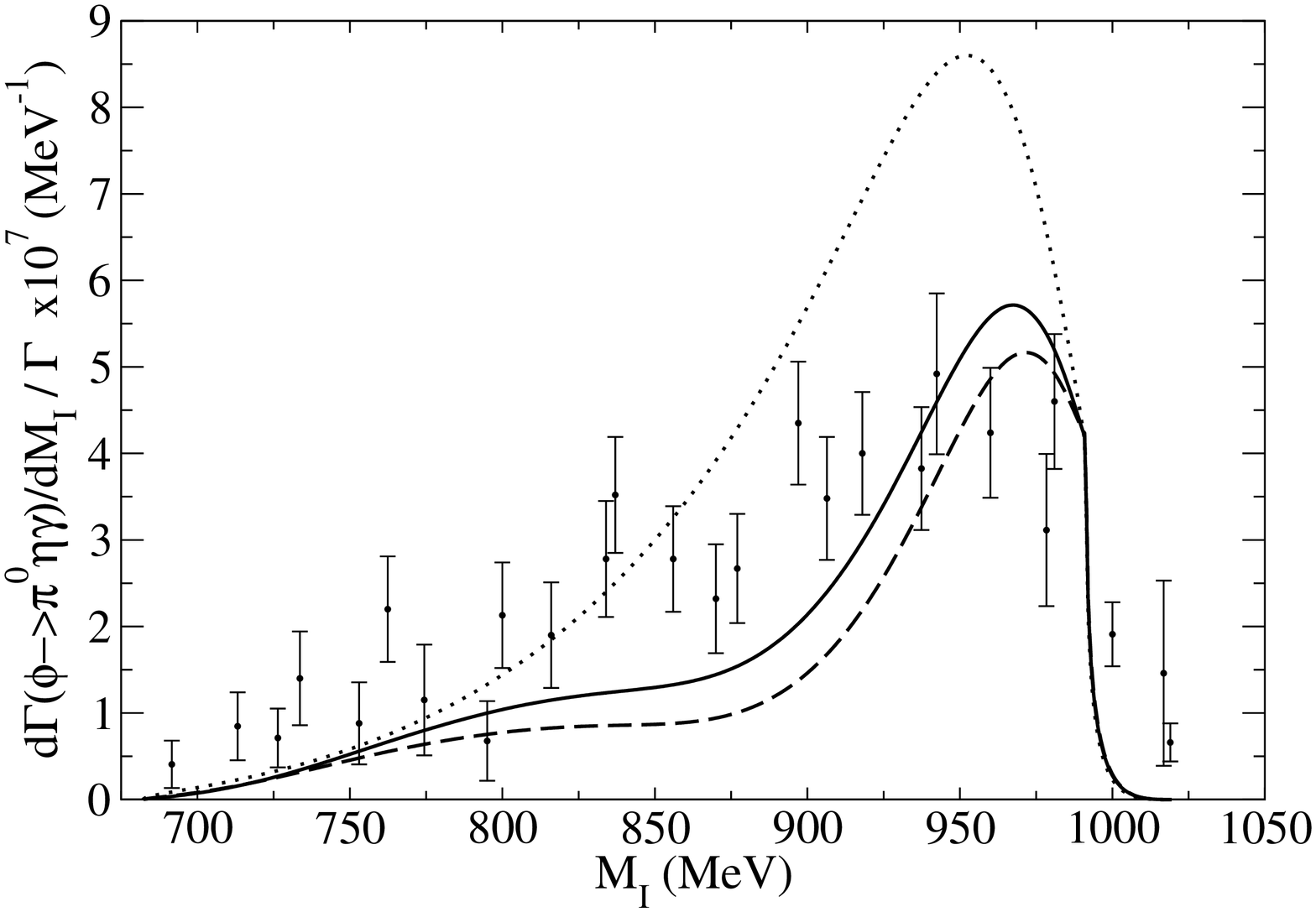,width=9cm,angle=-90}}}
\caption{\rm 
Different contributions to the $\pi^0\eta$ invariant mass
distributions of the $\phi\to\pi^0\eta\gamma$ decay:
Dotted line: chiral loops of Fig.~\ref{looplot}.
Dashed line: chiral loops of Fig.~\ref{looplot} + 
sequential VMD and its final
state interaction.
Solid line: idem plus the contribution of the mechanisms
involving axial-vector mesons, (full model).
}
\label{fig:res11}
\end{figure}

In Fig.~\ref{fig:res11} we show the different relevant contributions when
they are added one by one.

The dotted line represents the chiral loops by themselves. In
the dashed line we have added the contribution of the
sequential VMD mechanisms together with the loops involving
the FSI of the mesons. Finally, in solid line we have add
the contribution of the axial-vector contributions. This
latter curve represents the full model. In this figure we can
see the importance of the loop mechanism involving kaons and
the nonnegligible contribution of the mechanisms with
axial-vector mesons in intermediate states.

\begin{figure}[tbp]
\centerline{\hbox{\psfig{file=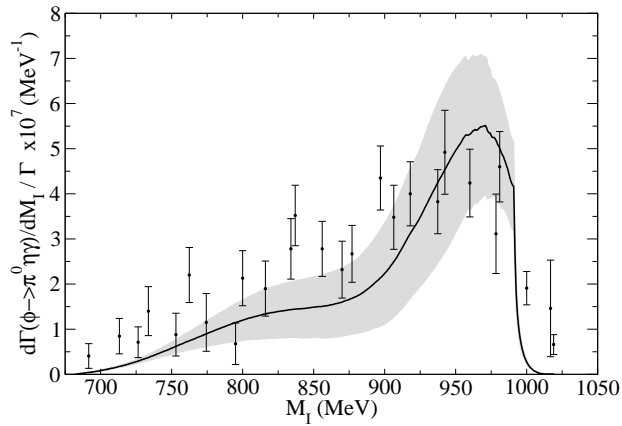,width=9cm,angle=-90}}}
\caption{\rm 
Final results for the $\pi^0\eta$ invariant mass distribution
for the $\phi\to\pi^0\eta\gamma$ decay with the theoretical
error band. }
\label{fig:res12}
\end{figure}

Again, in Fig.~\ref{fig:res12} we have plotted the
full model performing the
theoretical error analysis\footnote{
We have also checked that the use of a mixing angle for the
strange members of the axial nonets of around $30$ or $60$
degrees \cite{suzuki,axial} turns out in decreasing the lower
limit of the error band in around $5\%$ and $10\%$ for the 
$\phi\to\pi^0\pi^0\gamma$ and $\phi\to\pi^0\eta\gamma$ decays
respectively.
}.
We can see that when these   uncertainties are considered we
obtain a theoretical band
 in acceptable agreement with the
experimental data.

The total width and branching ratio obtained are

\begin{eqnarray}
\Gamma_{\phi\to\pi^0\eta\gamma}&=&250\pm 80 \textrm{ eV}\\
\nonumber
BR(\phi\to\pi^0\eta\gamma)&=&(0.59\pm 0.19)\times 10^{-4}
\end{eqnarray}
to be compared with the experimental values

\begin{eqnarray} \nonumber
BR^{exp}(\phi\to\pi^0\eta\gamma)
&=&(0.88\pm 0.14\pm 0.09)\times 10^{-4} 
\quad \textrm{ \cite{novo3}}\\
 \nonumber
BR^{exp}(\phi\to\pi^0\eta\gamma)
&=&(0.90\pm 0.24\pm 0.10)\times 10^{-4} 
\quad \textrm{ \cite{novo2}}\\
 \nonumber
BR^{exp}(\phi\to\pi^0\eta\gamma)
&=&(0.85\pm 0.05\pm 0.06)\times 10^{-4}
\quad \textrm{ \cite{frascati2}}
\end{eqnarray}

\section{
Further discussion of the $\phi\to\pi^0\pi^0\gamma$ results.
\label{sectionManolo}}

We would like to comment on the strength that we obtain
around $500\textrm{ MeV}$ in the $\pi^0\pi^0$
 invariant mass distribution
in the $\phi\to\pi^0\pi^0\gamma$ decay which appears in contradiction with the experimental
analysis. As we saw it comes from accumulation of the novel
mechanisms which we have discussed in our paper. Such
mechanisms are not considered in other theoretical papers
which find a good agreement with the data. For instance in 
\cite{lucio} a good reproduction of the data is obtained
using a  linear sigma model with $\sigma(500)$ and $f_0(980)$
included in the calculation and some free parameter to adjust
to the data. In \cite{pennington} an attempt to make a model
independent calculation is done by parametrizing amplitudes
compatible with analyticity and gauge invariance and, through
a fit to data, a set of parameters leading to agreement with
the data is obtained. In \cite{oller2} the same approach for
the chiral kaon loops from $\phi\to K^+K^-$ decay as in the
present paper is followed, but in addition a contact term in
$\phi\gamma K^0\bar{K^0}$ is added and, fitting the strength
of this term to the data, a good agreement with the
experimental results is also found. Our philosophy has been
different and we have not fitted any parameter to the $\phi$
radiative decay data but simple have considered the different
mechanisms that can sizeable contribute to the process. An
acceptable agreement with the data is found in the region of
the $f_0(980)$, which is the most important issue concerning
this reaction. This is not trivial a priori in view of the
very small width of the dynamically generated $f_0(980)$
(around $30\textrm{ MeV}$) that one obtains in the model of
\cite{npa} that we use here and the large 'visual' width of
the $f_0(980)$ peak in the present experiment. Part of the
reason for the agreement comes from the $q$ factor (photon
momentum) in the amplitude, as requirement by gauge
invariance and emphasize in \cite{achasovq,pennington}, which
gives more weight to the amplitude as we move down in the
$\pi\pi$ invariant mass from the upper limit (where $q=0$).
However, as seen in our results, the inclusion of the new
mechanisms and their interference with the dominant one,
particularly the contribution of the axial-vector meson
exchange mechanisms, also contributes to the widening of the
distribution around the $f_0(980)$ peak.

Although the agreement with  data at low masses is not very
good, we must point out two sources of uncertainty in the
experimental spectrum. First,  the results in the low and
intermediate mass region largely depend on  the background
subtraction dominated by the non-resonant $\omega\pi^0$ 
process. The size of this process is difficult to obtain
because  it has a strong background itself, mostly from the
$\phi\to f_0 \gamma$ process, as it is discussed in
\cite{kloe1}. There, its magnitude  has been obtained  in a
model dependent way assuming some a priori spectrum for the 
$\phi\to f_0 \gamma$ process \cite{kloe1}. In fact, before the
subtraction,  the raw data resemble much more our calculated
spectrum, (see fig. 4 from \cite{frascati1}), and we could
think of a slightly smaller $\omega\pi^0$ background.

Additionally, there is some  uncertainty in the way the data
are corrected to account for the experimental efficiency.
This is done in \cite{frascati1} by dividing the observed spectrum
by the effect of applying  the experimental efficiency on
some theoretical distribution. This unfolding
procedure depends on the theoretical model used, which  we 
think at low $\pi^0\pi^0$ masses is at least incomplete. 
 In fact, with the unfolding method used, the zero value 
of the spectrum obtained with the theoretical model of 
\cite{frascati1} implies unavoidably a zero value for the
corrected experimental results. A reanalysis to the light of the 
present discussion would be most welcome.


\section{Conclusions}
We have studied the radiative decay of the $\phi$ into two
neutral pseudoscalars, $\pi^0\pi^0$, $\pi^0\eta$, and a
photon. We have taken an approach to study the process
addressing mechanisms proved to be relevant in the study of
the radiative decays of the $\rho$ and the $\omega$. These
include the kaon loops from the $\phi\to K^+K^-$ decay and
mechanisms of sequential vector meson steps with vertices of
the type $VVP$, together with vector meson dominance to couple
a vector meson to the photon.

In addition to these mechanisms studied before we have added
the final state interaction in the sequential vector meson
mechanisms, allowing for rescattering of the $\pi^0\pi^0$ or
$\pi^0\eta$ states. Simultaneously we also allowed kaons to be
produced in the sequential vector meson processes, followed
by the interaction of the kaons to lead to the final
$\pi^0\pi^0$ or $\pi^0\eta$ states. These latter mechanisms
have the advantage over the tree level sequential vector meson
mechanism that they are not OZI forbidden.
We found these kaon loop contributions to be individually
large, although they become small in the
$\phi\to\pi^0\pi^0\gamma$ decay
 due to an accidental cancellation of
two terms, and they are very important
in the $\phi\to\pi^0\eta\gamma$ decay.

Another addition in the present work is the contribution of
sequential axial-vector mechanisms involving two sequential
$AVP$ steps followed by the coupling of a vector meson to the
photon, and even more important the production of
$K\overline{K}$  through these mechanisms followed by final
state interaction of the kaons to give $\pi^0\pi^0$ or
$\pi^0\eta$ in the final state.

The final state interaction of pairs of mesons involving
diagrams with loops has been done using techniques of
unitarized chiral perturbation theory which allow one to study
meson meson interaction in coupled channels
up to $1.2\,$GeV
and hence are particularly suitable for the present reaction.

The main contribution to the radiative $\phi$ decay is the
loop mechanisms involving the interaction of $K^+K^-$ coming
from $\phi$ decay, which by itself reproduces qualitatively
the experiment as it was already shown in \cite{uge}. However,
the new mechanisms studied here are by no means negligible.
Altogether we find a good
agreement with experiment, particularly when the study of the
theoretical uncertainties is done.
These
theoretical uncertainties, stemming from errors in the
magnitudes used as input to determine the parameters of the
theory, are of the same order of magnitude as the experimental
ones. Apart from these uncertainties we
should stress that there is no freedom in the theory used.
There are no free parameters which have been adjusted to the
$\phi\to\pi^0\pi^0\gamma$ and $\phi\to\pi^0\eta\gamma$ data
studied here.
The agreement with the experimental data both for
$\pi^0\pi^0$ and $\pi^0\eta$ channel is remarkable 
and less
than trivial when one realizes that a change of sign in one
of the mechanisms leads to opposite interference effects and
to results in not so good agreement with the data.
The agreement is more remarkable, or surprising, when one
realizes that this agreement with the data is reached for the
$\pi^0\pi^0\gamma$ case, in spite of having an $f_0(980)$
resonance that has a width much smaller than the apparent
width that one could guess
 from the $\phi\to\pi^0\pi^0\gamma$
 experiment, and which one would get
fitting the data with models less elaborate than the present
one. 
Some explanation for this can be already seen in 
\cite{achasovq,pennington} which remarked that the requirement
of gauge invariance in the amplitude introduces the factor of
the photon momentum which tends to widen the distribution of
the invariant mass around the $f_0(980)$ peak. Our approach is
manifestly gauge invariant and one  can show explicitly in
all the terms that this factor appears. Our approach comes to
stress the point of view of \cite{achasovq,pennington} that
the apparent width of the $f_0(980)$ in the experimental data
of this reaction can not be used as an direct measure of the
actual $f_0(980)$ width.

The main conclusion of the work is that a theoretical approach
to the problem
of the $\phi\to\pi^0\pi^0\gamma$ and $\phi\to\pi^0\eta\gamma$
decays is possible by 
using the chiral unitary approach used successfully
to reproduce the $\pi\pi$ data up to $1.2\,$GeV and
many other physical processes \cite{report}. As we could see
along the work, the $f_0(980)$ or $a_0(980)$ have not been
explicitly coupled to the $\phi$, meaning that there is no
direct $\phi\to f_0\gamma,a_0\gamma$ coupling. This is in
accordance with the philosophy of the chiral unitary approach
about these two resonances which are dynamically generated by
the meson meson interaction. This has as a consequence that in
processes where these resonances are produced one should not
include them explicitly in the approach but they should
appear naturally as soon as the meson are allowed to interact.
In this sense the present work comes to reinforce the claim
that the  $f_0(980)$ and $a_0(980)$ are dynamically generated
resonances and that the meson meson chiral Lagrangian at
lowest order contains the driving forces that make possible
this generation through the multiple scattering of the mesons
in coupled channels.

\section*{Acknowledgments}
Two of us, J.E.P. and L.R., acknowledge support from the
Ministerio de Educaci\'on, Cultura y Deporte. 
This work is
partly supported by DGICYT contract number BFM2000-1326,
and the E.U. EURIDICE network contract no. HPRN-CT-2002-00311.

\vspace{2cm}

{\bf \Large{Appendix:
Integrals involving three meson propagators}}

\begin{figure}
\centerline{\protect\hbox{
\psfig{file=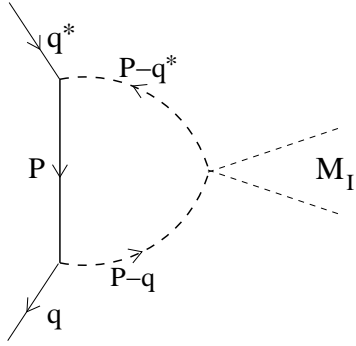,width=0.3\textwidth,silent=}}}
\caption{Three meson loop. }
\label{fig:loop_3meson}
\end{figure}

For the evaluation of the loop of Fig.~\ref{fig:loop_3meson}
we need to calculate
the integrals of Eq.~(\ref{eq:Is}).
From Eq.~(\ref{eq:Is}) it is easy to obtain that

\begin{eqnarray} \label{eq:Isb}
I_0&=&i\int\frac{d^4P}{(2\pi)^4} P^0P^0 D_1 D_2 D_3
\\ \nonumber
I_1&=&i\int\frac{d^4P}{(2\pi)^4} P^0 |\vec{P}|
\cos{\theta} D_1 D_2 D_3 \\
\nonumber
I_a&=&\frac{i}{2}\int\frac{d^4P}{(2\pi)^4} \vec{P}\,^2 
\sin^2{\theta} D_1 D_2 D_3 \\
\nonumber
I_b&=&\frac{i}{2}\int\frac{d^4P}{(2\pi)^4} \vec{P}\,^2 
(3\cos^2{\theta}-1) D_1 D_2 D_3
\end{eqnarray}

where
$D_1=\frac{1}{P^2-M_V^2+i\epsilon}$, 
$D_2=\frac{1}{(P-q^*)^2-m_1^2+i\epsilon}$ and
$D_3=\,\frac{1}{(P-q)^2-m_2^2+i\epsilon}$.

\begin{figure}
\centerline{\protect\hbox{
\psfig{file=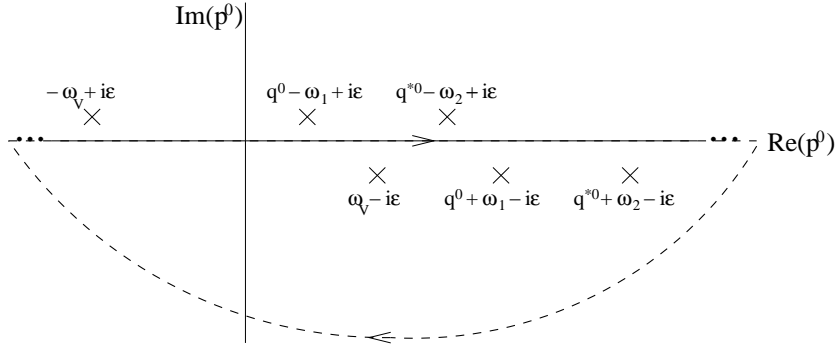,width=0.7\textwidth,silent=}}}
\caption{Poles and path for the evaluation
of the $P^0$ integral.}
\label{fig:poles}
\end{figure}

Defining 
$\omega_V=\sqrt{\vec{P}\,^2+M_V^2}$, 
$\omega_2=\sqrt{{(\vec{P}-\vec{q})}\,^2+m_2^2}$ and
$\omega_1=\sqrt{{(\vec{P}-\vec{q}^*)}^2+m_1^2}$,
the analytical structure in the complex $P^0$ plane of the
integrals of Eqs.~(\ref{eq:Isb}) is depicted in
 Fig.~\ref{fig:poles}.

Applying the Residues Theorem with the path shown
in Fig.~\ref{fig:poles},
the evaluation of the $P^0$ integral gives 

\begin{eqnarray} \label{eq:IsinP0}
\nonumber
I_i=\frac{1}{2}\int\frac{d\cos\theta\,d|\vec{P}|}{(2\pi)^2}
\frac{|\vec{P}|^2}{\omega_1\omega_2}&& \frac{\tilde{A_i}}
{(q^{*0}+\omega_1+\omega_V)
 (q^{*0}-\omega_1-\omega_V+i\frac{\Gamma_V(s'_V)}{2})
 (q^{*0}-q^0+\omega_1+\omega_2-i\epsilon)
 } \cdot \\
 &\cdot&\frac{1}{
 (q^{0}+\omega_2+\omega_V)
 (q^0-\omega_2-\omega_V+i\frac{\Gamma_V(s_V)}{2})
 (q^0-q^{*0}+\omega_1+\omega_2-i\epsilon)
 }
\end{eqnarray}

for $i=\{0,1,a,b\}$, 
with $s_V=(q^{*0}-\omega_1)^2-\vec{P}\,^2$, 
$s'_V=(q^{0}-\omega_2)^2-\vec{P}\,^2$ and

\begin{eqnarray} \label{eq:Agorros}
\nonumber
\tilde{A}_0&=&2q^0q^{*0}\omega_1 \omega_2\omega_V
-\omega_2(\omega_2+\omega_V)\left[\omega_1
(\omega_1+\omega_2)(\omega_1+\omega_V)
-{q^{*0}}^2(\omega_1+\omega_2+\omega_V)\right] +
\\ \nonumber
&+&{q^0}^2\left[-{q^{*0}}^2(\omega_1+\omega_2)
+\omega_1(\omega_1+\omega_V)
(\omega_1+\omega_2+\omega_V)\right] \\
\nonumber
\tilde{A}_1&=&|\vec{P}|\cos\theta
\left\{{-{q^0}^2q^{*0}}^2\omega_2
+q^{*0}\omega_2(\omega_2+\omega_V)(2\omega_1+\omega_2+w_V)+
\right.
\\ \nonumber
&+&\left.{q^0}\omega_1\left[-{q^{*0}}^2+(\omega_1+\omega_V)
(\omega_1+2\omega_2+\omega_V)\right]\right\}
\\ \nonumber
\tilde{A}_a&=&\frac{\vec{P}\,^2 
\sin^2{\theta}}{2\omega_V}
\left\{ 2q^0q^{*0}\omega_1\omega_2-{q^0}^2\omega_2
(\omega_1+\omega_V) + \right. \\
 \nonumber
 &+&\left.(\omega_2+\omega_V)\left[-{q^{*0}}^2\omega_1+
(\omega_1+\omega_2)
(\omega_1+\omega_V)(\omega_1+\omega_2+\omega_V)\right]
\right\}
\\ \nonumber
\tilde{A}_b&=&\tilde{A}_a\frac{
(3\cos^2{\theta}-1)}{\sin^2{\theta}}
\end{eqnarray}

\end{document}